\documentclass[10pt,aps,prd,noshowpacs,eqsecnum,nofootinbib,twocolumn]{revtex4-1}
\usepackage{amsmath,amsthm,amssymb}
\usepackage{mathrsfs,bbm}

\usepackage{nicefrac} 

\usepackage{longtable}

\usepackage{latexsym}

\usepackage{array}

\usepackage{color}

\begin{document}


\title{Reply to ``Comment on `Texture Zeros and WB
Transformations in the Quark Sector of the
Standard Model''' }



\author{Yithsbey Giraldo}
\affiliation{Departamento de F\'\i sica, Universidad de Nari\~no, A.A. 1175, San Juan de Pasto, Colombia}
\date{\today}

\begin{abstract}
We respond to the comments of S. Sharma et al.~[Phys. Rev. D 91, 038301 (2015)] on my recent paper, Y. Giraldo~[Phys. Rev. D 86, 093021 (2012)]. In their comments, they begin discussing a particular class of five-zero texture non-Fritzsch-like quark mass matrix, which was proposed by me, and questioning its validity. Then, they put in doubt the phases included in the unitary matrices used for diagonalizing the quark mass matrices, from which they claim that the CP violating parameter $\epsilon_k$ obtained does not agree with the experimental value. Because of these comments, finally, they recommend being careful while analyzing the implications of Weak Basis transformations on textures.  Other minor points are also discussed by them.

In the following,  I will show that the mentioned five-zero texture non-Fritzsch-like quark mass matrices is completely valid and generates all the physical quantities involved,  including the CP violating parameter $\epsilon_k$, for which is required the inclusion of phases in the unitary matrices used for diagonalizing the quark mass matrices in order to put the CKM matrix in standard form. These phases can be rotated away so they do not have any physical meaning.  Finally, the relevance of the weak basis transformation is appreciated: which is  {\it complete} and facilitates calculations, especially of textures zeros. 

\end{abstract}


\maketitle


\section{Introduction}
In Standard Model (SM), the most general weak basis~(WB) transformation~\cite{b2}, that leaves the physical content invariable and the up- and down-quark mass matrices  $M_{u}$ and $M_{d}$ Hermitian~\footnote{The quark mass matrices are Hermitian due to the polar decomposition theorem, where the unitary component can be absorbed in the right-handed quark fields.}, is
\begin{equation}
\label{1.1a}
 \begin{split}
  M_u&\longrightarrow M_u^\prime=U^\dag M_u U,\\
M_d&\longrightarrow M_d^\prime=U^\dag M_d U,
 \end{split}
\end{equation}
where $U$ is an arbitrary unitary matrix. We say that the two quark mass matrices $M_{u,d}$ and $M_{u,d}^\prime$ are equivalent each other.  So, this implies that the number of equivalent mass matrices is  infinity. 
Hence, we are able to explicitly construct texture  zeros in quark mass matrices through WB transformations. If these texture zeros exist, the WB transformation is able to find them.  The reason is that, as was shown in my paper~\cite{b1}, the WB transformation is exhaustive~(complete) finding all possibilities, included possible four and five-zero textures. 

Through WB transformations, Branco et al.~\cite{b2}  show that is always possible to find, at most, three zeros in quark mass matrices with no physical meanings. But, this does not restrict the number of zeros can be found by applying the WB transformation to mass matrices,  the case is that the model must be put into a physical context. Therefore, we have found  additional zeros~{(four- and even five-zero textures~\cite{b1})} by using the recent quark mass and mixing data. These additional zeros now have physical meanings because they were obtained from specific experimental data. 
{\section{up-quark Mass Matrix in Diagonal Form}}
One point of discussion is that to facilitate the analysis the initial quark mass matrices used by me is as follows~\cite{b1}

\begin{equation}
\label{1.1}
\begin{split}
 M_u&=D_u=\begin{pmatrix}
          \lambda_{1u}&0&0\\
0&\lambda_{2u}&0\\
0&0&\lambda_{3u}
         \end{pmatrix},\\
%
 M_d&=VD_dV^\dag,
\end{split}
\end{equation}
where the up and down diagonal matrices $D_u$ and $D_d$ contain the respective quark mass eigenvalues, and $V$ is the usual  quark  Cabibbo-Kobayashi-Maskawa~(CKM) mixing  matrix. The authors' comments claim that we do not start with the most general mass matrices. But this is not true.  The starting matrices~\eqref{1.1}, used in papers like~\cite{b2,b3},  are as general as any other {one}. The reason is that starting from arbitrary Hermitian matrices $M_u$ and $M_d$, and using their respective diagonalizing matrices  $U_u$ and $U_d$,  and performing a  WB transformation~\eqref{1.1a} using for this case  the unitary matrix $U=U_u$. We have
\begin{equation*}
 \begin{split}
  M_u\longrightarrow M'_u&=U^\dag_u\, M_u\, U_u=D_u,
\\
M_d=U_d\,D_d \,U_d^\dag\longrightarrow M'_d&=U^\dag_u\,(U_d \,D_d \,U_d^\dag)\,U_u,
\\
&=(U^\dag_u\,U_d)\,D_d\,(U_d^\dag \,U_u),
\\
&=V \,D_d \,V^\dag ,
 \end{split}
\end{equation*}
where the CKM mixing matrix $V=U^\dag_u\, U_d$ was considered.  Additionally, note also that the three no-physical-texture zeros mentioned above appear also in~\eqref{1.1}.

Although, the crux  of the comments is below. 

\section{Phases and The CMK Mixing Matrix}
Let us resolve the problem for a particular case.  Let us consider the numerical quark mass matrices given in Eq.~(4.22) of my paper~\cite{b1}. Which was also considered by the authors' comments in  row (a) of Table 1.  Apparently, the corresponding CKM matrix obtained is not compatible with the recent quark mixing data. The numerical quark mass matrices  in discussion are~(in MeV units)
\begin{widetext}
{\small
\begin{align*}
M_u&=\begin{pmatrix}
 0 & 0 & -{92.3618}+{157.694} i \\
 0 & {5748.17} & {28555.1}+{5911.83} i \\
 -{92.3618}-{157.694} i & {28555.1}-{5911.83} i &
   166988
\end{pmatrix},
\quad
M_d&=\begin{pmatrix}
 0 & {13.9899} & 0 \\
 {13.9899} & 0 & {424.808} \\
 0 & {424.808} &{2796.9}
\end{pmatrix},
\end{align*}}
%
where their diagonalizing matrices are  respectively\footnote{These diagonalizing  matrices were obtained by using Mathematica, but different matrices are obtained by using other software like maxima, octave, \ldots. Actually, the difference is just in the phases.}
%
{\footnotesize
\begin{equation}
\label{2.1}
U_u=\begin{pmatrix}
0.998779\times e^{2.10064 i} & 0.0493829\times e^{2.10064 i} &
   0.00104594\times e^{2.10064 i} \\
 0.0484608\times e^{0.20415i} & 0.983788\times e^{-2.93744i} & 0.172662\times e^{0.204148 i} \\
 -0.00955555\times e^{0i} & 0.172401\times e^{0i} &
   0.984981\times e^{0i} 
\end{pmatrix},
\quad
U_d=\begin{pmatrix}
{0.978718} & -{0.205210} &
   {0.000718698} \\
 {0.202880} & {0.968118} & {0.146926} \\
 -{0.0308464} & -{0.143653} &
   {0.989147} 
\end{pmatrix},
\end{equation}}
%
from which the authors' comments have obtained the quark mixing matrix, Eq.~(7). Nevertheless, generalizing the result in~\eqref{2.1},  no physical  phases~($x$, $y$, $z$, $v$ and $w$) can be added to the matrices,  obtaining the following diagonalizing matrices 
%
\begin{subequations}
\label{2.2}
{\small
\begin{align}
U'_u=&\begin{pmatrix}
0.998779\times e^{2.10064 i}\times e^{x i} & 0.0493829\times e^{2.10064 i}\times e^{y i} &
   0.00104594\times e^{2.10064 i} \times e^{z i}
\\
 0.0484608\times e^{0.20415i}\times e^{x i} & 0.983788\times e^{-2.93744i}\times e^{y i} & 0.172662\times e^{0.204148 i} \times e^{z i}
\\
 -0.00955555\times e^{0i}\times e^{x i} & 0.172401\times e^{0i}\times e^{y i} &
   0.984981\times e^{0i} \times e^{z i}
\end{pmatrix},
\\
U'_d=&\begin{pmatrix}
{0.978718}\times e^{iv} & -{0.205210}\times e^{iw} &
   {0.000718698} \\
 {0.202880}\times e^{iv} & {0.968118}\times e^{iw} & {0.146926} \\
 -{0.0308464}\times e^{iv} & -{0.143653}\times e^{iw} &
   {0.989147} 
\end{pmatrix},
\end{align}}
\end{subequations}
%
and where there is no way to distinguish what are the ``true'' matrices of diagonalization.
Even further, if you choose the values $x=-1.30524$, $y=0.790611$, $z=-0.00515513$, $v=0.785572$ and $w=-2.14216$, you obtain 
{\small
\begin{align*}
U'_u=&\begin{pmatrix}
0.998779\times e^{0.795395 i} & 0.0493829\times e^{2.89125 i} &
  0.00104594\times e^{2.09548 i} 
\\
 0.0484608\times e^{-1.10109 i} &0.983788\times e^{-2.14683 i} & 0.172662\times e^{0.198993 i} 
\\
 0.00955555\times e^{1.83635 i} & 0.172401\times e^{0.790611 i} &
 0.984981\times e^{-0.00515513 i} 
\end{pmatrix},
\\
U'_d=&\begin{pmatrix}
{0.978718}\times e^{0.785572 i} & -{0.205210}\times e^{-2.14216 i} &
   {0.000718698} \\
 {0.202880}\times e^{0.785572 i} & {0.968118}\times e^{-2.14216 i} & {0.146926} \\
 -{0.0308464}\times e^{0.785572 i} & -{0.143653}\times e^{-2.14216 i} &
   {0.989147} 
\end{pmatrix},
\end{align*}}
%
two diagonalizing matrices that are equally valid. As a result, a quark CKM  mixing matrix compatible with the recent mixing data~\cite{b6} is derived
{\small
\begin{equation}
\label{3.3}
V_{ckm}=U_u^{\prime\dag}\cdot U'_d= 
\begin{pmatrix}
 0.974276& 0.225334 &0.00124462 - 0.0032841 i
\\
-0.225194-0.000106564 i& 0.973443-0.0000294788 i& 0.0411845 
\\
0.00806881 - 0.00319789 i&-0.0404056 - 0.000739786 i& 0.999145
 \end{pmatrix}.
\end{equation}}
\end{widetext}
As you can observe, the phases included in~\eqref{2.2} are not against to reduce the number of free parameters. These phases come out naturally from the diagonalizing process, and  it is impossible to avoid them. When you establish specific diagonalization matrices, you are choosing specific phases.
These phases  are just different ways to present the CKM matrix as was told in my paper~\cite{b1} above equation (3.29). 

Going even further these phases have an interpretation  and its nature is clarified in the next section.
\section{No Physical Phases}
The additional phases introduced  in matrices~\eqref{2.2} leave the physical content invariable, including the Jarlskog invariant quantity. This can be seen by bringing the matrices~\eqref{2.2} to the  following products
\[
 U'_u= U_u\,f_1\quad\textrm{and}\quad U'_d= U_d\,f_2,
\]
 where the diagonal matrices $f_1=\textrm{diag}(e^{xi},e^{yi},e^{zi})$ and $f_2=\textrm{diag}(e^{vi},e^{wi},1)$; and $U_u$ and $U_d$ are given in~\eqref{2.1}. Such that the CKM mixing matrix obtained in~\eqref{3.3} becomes
\[
 V_{ckm}=U_u^{\prime\dag}\,U'_d=(U_u\,f_1)^\dag\,(U_d\,f_2)=f_1^\dag\,(U_u^\dag\,U_d)\,f_2,
\]
therefore
\begin{equation}
\label{2.3}
 U_u^\dag\,U_d= f_1\,V_{ckm}\,f_2^\dag,
\end{equation}

which implies the following two results:
\begin{enumerate}
 \item First, the mixing matrix obtained from Eq.~\eqref{2.1}, i.e.  $U_u^\dag\,U_d$, apparently does not fit the standard form of the CKM mixing matrix, but  is well known that the five phases present in $f_1$ and $f_2$ can be rotated away~\cite{b4}, such that both expressions for the CKM, in~\eqref{2.3},   are equivalent. Therefore the Jarlskog invariant, as well as any other physical quantity, is not affected by adding phases as given in~\eqref{2.2}. And as a result, the phases $x,y,z,v$ and $w$ have no physical meaning.  
\item Second, if there are still doubts, $f_1$ and $f_2$ in~\eqref{2.3} add phases to the matrix elements of $V_{ckm}$, where each phase of $f_2$  is placed in the same column and each phase of $f_1$ in the same row. Therefore, the Jarlskog invariant $J=\textrm{Im}(V_{us}V_{ub}^*V_{cs}^*V_{cb})$ is not affected by adding these additional phases, because they cancel out in the same row $(V_{us}, V_{ub}^*)$ and $(V_{cs}^*, V_{cb})$,  and in the same column $(V_{us}$, $V_{cs}^*)$ and  $(V_{ub}^*, V_{cb})$.
\end{enumerate}
Finally, the Jarlskog invariant quantity is
\[
 J=\textrm{Im}(V_{us}V_{ub}^*V_{cs}^*V_{cb})=2.96695\times10^{-5},
\]
clearly inside the range given by PDG 2012~\cite{b6}, i.e., $(2.80-3.16)\times10^{-5}$. The same for the quark masses~(in MeV): $m_d=2.90, m_s=66, m_b=2860$, $m_u=1.75, m_c=638, m_t=172100$~\cite{b6}.

We can consider other phase invariant quantities  like the inner angles of the CKM unitarity triangle:
$\beta=\arg(-\mbox{\footnotesize ${V_{cd}V_{cb}^*}/{V_{td}V_{tb}^*})$}=21.6^\circ$,  
$\alpha=\arg(-\mbox{\footnotesize ${V_{td}V_{tb}^*}/{V_{ud}V_{ub}^*})$}=89.1^\circ$, and $\gamma=\arg(-\mbox{\footnotesize ${V_{ud}V_{ub}^*}/{V_{cd}V_{cb}^*})=69.2^\circ$}$.  

However, in the frame of the SM, the usual formula for the Kaon CP violating parameter $\epsilon_k$, is valid only in the basis where $V_{ud}V_{us}^*$ is real{~\cite{b6,b5}}, for that reason the phases given in~\eqref{2.2} must be considered  in order to transform the CKM mixing matrix into its standard convention.

\section{Conclusions}

To begin with, {the WB transformation is complete, so we can find all possible quark mass matrices representing the model {by} starting from a specific quark mass matrices.}  It is important to mention that, in the SM, {is always possible to find a maximum of three no physical vanishing elements in the quark mass matrices by performing a WB transformation.} In the process does not matter the value of physical quantities. But if we want  to find additional zeroes is necessary to take into account physical considerations.

Another important result, emphasized by other authors, is that the quark mass matrix  structure given in~\eqref{1.1}, which was {called} in my  paper~\cite{b1} as the{ \it u-diagonal representation}, is so general as any other one. These matrices are deduced from a WB transformation and  have the advantage of having available the quark masses and the CKM matrix elements.

The phases included in the diagonalizing matrices~\eqref{2.2}, are precisely the five phases that can be rotated away through the phase redefinition of the left-handed up and down quark fields as was shown in~\eqref{2.3}, and as a consequence these phases have no physical meaning. Even further, as a result, these phases does not affect the invariance of the Jarlskog quantity, and so on. Respect to the Kaon CP violating parameter $\epsilon_k$, it must be calculated in a basis where $V_{ud}V_{us}^*$ is real. For that reason, phases must be included in the diagonalizing matrices in order to achieve this requirement. 

Definitely, the introduction of extra additional phases is not against the basic spirit of the texture specific mass matrices which was to control the number of free
parameters. These phases have no physical meaning, but their inclusion is necessary to adjust the resulting CKM matrix to its standard choice.


\section*{Acknowledgments}
This work was partially supported by Department of Physics in the Universisad de Nari\~no, approval Agreement Number 009.


\begin{thebibliography}{99}
\bibitem{b2} G. C. Branco, D. Emmanuel-Costa and R. G. Felipe, Phys. Lett. B 477, 147
(2000) [hep-ph/9911418]; D. Emmanuel-Costa and C. Simoes, Phys. Rev. D 79, 073006 (2009) [ arXiv:0903.0564 [hep-ph]].

\bibitem{b1} Y. Giraldo, Phys. Rev. D 86, 093021 (2012), [arXiv:1110.5986 [hep-ph]].

\bibitem{b3} H. Fusaoka and Y. Koide, Phys. Rev. D 57, 3986 (1998).

\bibitem{b4} Andrija Rasin, [hep-ph/9708216].

\bibitem{b6} J. Beringer et al., Particle Data Droup, Phys. Rev. D 86, 010001 (2012).

\bibitem{b5} G. Buchalla {\it et al.}, Rev. Mod. Phys. 68, 1125 (1996) [hep-ph/9512380].



\end{thebibliography}
\end{document}